\documentclass[manuscript,nonacm]{CHI-ACM-contents/acmart}
\AtBeginDocument{%
  }

\setcopyright{acmlicensed}
\copyrightyear{2024}
\acmYear{2024}
\acmDOI{XXXXXXX.XXXXXXX}

\acmConference[CHI `25]{}{May 11--16,
  2024}{Honolulu, HI}
\acmISBN{978-1-4503-XXXX-X/18/06}

\acmSubmissionID{2857}

\usepackage{tabularx}
\usepackage{tabulary}
\usepackage{capt-of}

\usepackage{csquotes}

\usepackage{color, colortbl}

\usepackage{makecell}

\usepackage{booktabs,siunitx}

\usepackage{lipsum}

\usepackage{enumitem}

\usepackage{subfigure}

\usepackage{subcaption}
\usepackage{wrapfig}
\usepackage{xspace}
\usepackage[normalem]{ulem}
\definecolor{mred}{rgb}{.80,.12,.30}
\definecolor{MRED}{rgb}{.80,.12,.30}
\definecolor{grey}{rgb}{0.5,0.5,0.5}
\definecolor{purple}{rgb}{.75,0,.85}
\definecolor{pistachio}{rgb}{0.58, 0.77, 0.45}
\definecolor{palesilver}{rgb}{0.9, 0.9, 0.9}

\newcommand{\sys}{LinkQ\xspace}

\newif\ifnotes
\notestrue

\let\origcite\cite
\renewcommand{\cite}[1]{\ifnotes\mbox{\origcite{#1}}\else \origcite{#1}\fi}

\begin{document}

\title[Mitigating LLM Hallucinations with Knowledge Graphs: A Case Study]{Mitigating LLM Hallucinations with Knowledge Graphs: A Case Study}
% \title{A Preliminary Roadmap to Integrate LLMs and Knowledge Graphs for Visual Data Exploration}

% \author{
%     Harry Li$^{1}$,
%     Gabriel Appleby$^{2}$,
%     Kenneth Alperin$^{1}$,
%     Steven R Gomez$^{1}$,
%     Ashley Suh$^{1}$
%     \\
%     {\small $^1$MIT Lincoln Laboratory}
%     {\small $^2$National Renewable Energy Laboratory}
%     % {\vspace{-10pt}}
% }

% Ashley Suh
% MIT Lincoln Laboratory
% Lexington, Massachusetts, USA
% ashley.suh@ll.mit.edu

\author{Harry Li}
\email{harry.li@mit.ll.edu}
\affiliation{%
  \institution{MIT Lincoln Laboratory}
  \city{Lexington}
  \state{Massachusetts}
  \country{USA}
}

\author{Gabriel Appleby}
\email{Gabriel.Appleby@nrel.gov}
\affiliation{%
  \institution{National Renewable Energy Laboratory}
  \city{Golden}
  \state{Colorado}
  \country{USA}
}

\author{Kenneth Alperin}
\email{Kenneth.Alperin@mit.ll.edu}
\affiliation{%
  \institution{MIT Lincoln Laboratory}
  \city{Lexington}
  \state{Massachusetts}
  \country{USA}
}

\author{Steven R Gomez}
\email{Steven.Gomez@mit.ll.edu}
\affiliation{%
  \institution{MIT Lincoln Laboratory}
  \city{Lexington}
  \state{Massachusetts}
  \country{USA}
}

\author{Ashley Suh}
\email{Ashley.Suh@mit.ll.edu}
\affiliation{%
  \institution{MIT Lincoln Laboratory}
  \city{Lexington}
  \state{Massachusetts}
  \country{USA}
}

% \orcid{0000-0002-2288-6039}

% \author{Gabriel Appleby$^{2}$}
% \email{gabriel.appleby@tufts.edu}

% \author{Steven Gomez$^{1}$}
% \email{steven.gomez@ll.mit.edu}

% \author{Kenneth Alperin$^{1}$}
% \email{kenneth.alperin@ll.mit.edu}

% \author{Ashley Suh$^{1}$}
% \orcid{0000-0001-6513-8447}

% \email{ashley.suh@mit.ll.edu}
% \affiliation{%
%   \institution{MIT Lincoln Laboratory}
%   \streetaddress{244 Wood Street}
%   \city{Lexington}
%   \state{MA}
%   \postcode{02421}
% \affiliation{%

\renewcommand{\shortauthors}{Li et al.}

%% high stakes domains like defense, cyber, etc. needs trustworthy AI methods --> everyone wants to use LLMs because they're really helpful --> but they hallucinate --> we're going to show that using LLMs to query data for ground truth question + answering is better than not --> quantitative eval of gpt 4 vs linkq --> maybe some user quotes 
\begin{abstract}
High-stakes domains like cyber operations need responsible and trustworthy AI methods. While large language models (LLMs) are becoming increasingly popular in these domains, they still suffer from hallucinations. This research paper provides learning outcomes from a case study with LinkQ, an open-source natural language interface that was developed to combat hallucinations by forcing an LLM to query a knowledge graph (KG) for ground-truth data during question-answering (QA).
We conduct a quantitative evaluation of LinkQ using a well-known KGQA dataset, showing that the system outperforms GPT-4 but still struggles with certain question categories -- suggesting that alternative query construction strategies will need to be investigated in future LLM querying systems. We discuss a qualitative study of LinkQ with two domain experts using a real-world cybersecurity KG, outlining these experts' feedback, suggestions, perceived limitations, and future opportunities for systems like LinkQ.
% LinkQ is open-source at \url{https://github.com/mit-ll/linkq}.
\end{abstract}



\keywords{Large language models, knowledge graphs, visual analytics. \\\textbf{Presented at the Human-centered Explainable AI Workshop (HCXAI) @ CHI 2025.} DOI: \href{https://doi.org/10.5281/zenodo.15170463}{10.5281/zenodo.15170463}}

% Add the DOI information as a final keyword to your existing list of user-defined keywords:

\begin{teaserfigure}
  \centering
  \includegraphics[width=1\linewidth]{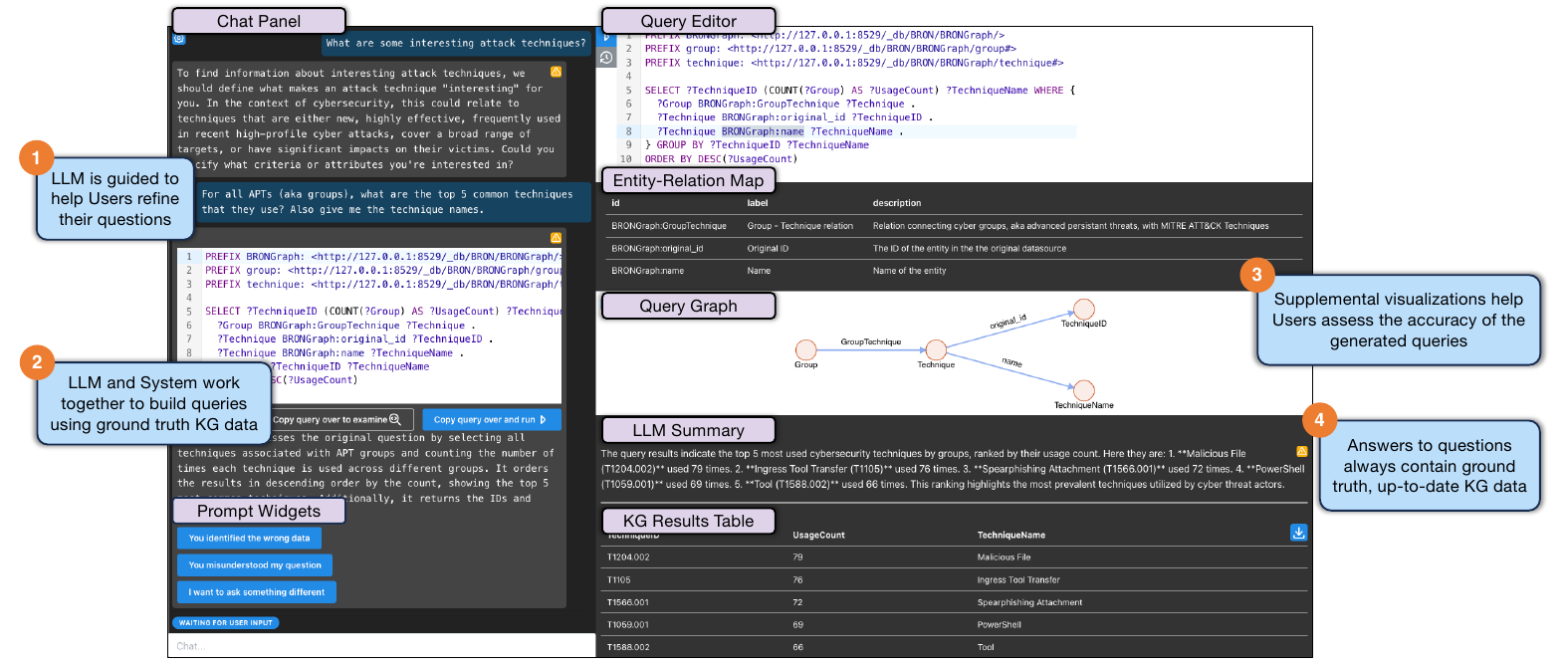}
  \caption{The \sys system~\cite{li2024linkq}, a natural language interface that combats an LLM's tendency to hallucinate by instructing it to answer users' questions by creating and executing knowledge graph (KG) queries. \sys ensures all data retrieved and summarized by the LLM comes from ground truth, up-to-date data in the KG.}
  \label{fig:teaser}
\end{teaserfigure}

\maketitle

% Content for the paper
\section{Introduction}

The development of trustworthy artificial intelligence (AI) methods is crucial for high-stakes domains to avoid unintended and potentially catastrophic consequences of using AI-enabled systems~\cite{doshi2017towards, amodei2016concrete}.
As large enterprises and government entities adopt Responsible AI policies~\cite{leslie2019understanding}, it is important for the research community to develop AI capabilities that help operators not just understand a model's outputs, but also provide them with the tools to interrogate and question the data associated with a model~\cite{krause2016interacting, wexler2019if}. 

Large language models (LLMs) have gained popularity in national security applications~\cite{motlagh2024large, caballero2024large, mikhailov2023optimizing}, partly for their ability to support question-answering over large amounts of text that operators cannot keep up with alone~\cite{tan2023can}.
However, LLMs still produce hallucinations, biases, and \textit{``intentional lies}''~\cite{li2024prelim}, which can lead to inaccurate, misleading results and ultimate distrust in AI~\cite{hofmann2024dialect}.
This workshop paper provides learning outcomes for this area of research through two new evaluations of the LinkQ system~\cite{li2024linkq}. LinkQ is a natural language interface system that combats an LLM's tendency to hallucinate during question-answering by forcing the LLM to answer users' questions by constructing and executing queries to a knowledge graph (KG). KGs are used to store complex, meaningful knowledge in graph form as a collection of nodes (\textit{entities}), edges (\textit{relations}), and properties (\textit{attributes}). Due to their robustness, KGs are a popular data structure for many real-world applications such as question-answering and recommendation~\cite{ehrlinger2016towards,wei2020combining, alkhamissi2022review, hogan2021knowledge, li2024kgs}.
% and common language tasks, particularly when combined with large language models (LLMs)~\cite{alkhamissi2022review, petroni2019language, brown2020language}. However, KGs are challenging to use in practice~\cite{li2024kgs, lissandrini2022knowledge, hogan2021knowledge, 2024_unifying_llms_and_kgs, lissandrini2020graph}, particularly when querying relevant insights. Even expert KG users may struggle to acquire data for downstream analyses due to the time-consuming and tedious process of crafting correct queries across multiple KG platforms.~\cite{li2024kgs}.

% In order to develop an AI-enabled question-answer system design with Responsible AI practices in mind, we developed \textbf{\sys}: a natural language interface system that utilizes an LLM to assist users in exploring and answering questions about ground-truth Knowledge Graph (KG) data~\cite{linkq}. Rather than having an LLM directly answer a user's question and risk hallucinations, LinkQ uses the LLM to instead generate a verifiable query, which is executed to retrieve ground-truth, up-to-date, trustworthy data directly from a KG. 
%For instance, in Figure~\ref{fig:teaser}, LinkQ's LLM (GPT-4) accurately commonly attack techniques used by cyber attackers, despite not being trained on that data.
We conducted a quantitative evaluation using a standardized KG question-answering (KGQA) dataset~\cite{sen2022mintaka} (Section~\ref{sec:results}), as well as a qualitative study with two domain experts using the BRON~\cite{hemberg2020linking} cybersecurity KG (Section~\ref{sec:demonstration}). Our quantitative evaluation shows that \sys strongly outperforms OpenAI's GPT-4 in generating accurate KG queries, but could not answer all question types with ideal accuracy. This suggests that, while LinkQ can be used for more trustworthy question-answering, future querying strategies will need to be implemented for better generalizability. Our qualitative study illuminates a real-world scenario of domain experts using an LLM to query a domain-specific KG, opening new directions of research in this area of LLM-based question-answering without hallucinations.
% \textit{without} any fine-tuning~\cite{zhang2023dissecting} or embeddings (i.e. a traditional retrieval augmented generation or RAG approach~\cite{rag}) -- methods that can be too narrow, costly, or difficult to scale~\cite{rangel2024sparql}
% \input{sections/related}
\section{LinkQ System Design}
\label{sec:design}

\begin{figure}[t]
    \centering
    \includegraphics[width=.99\linewidth]{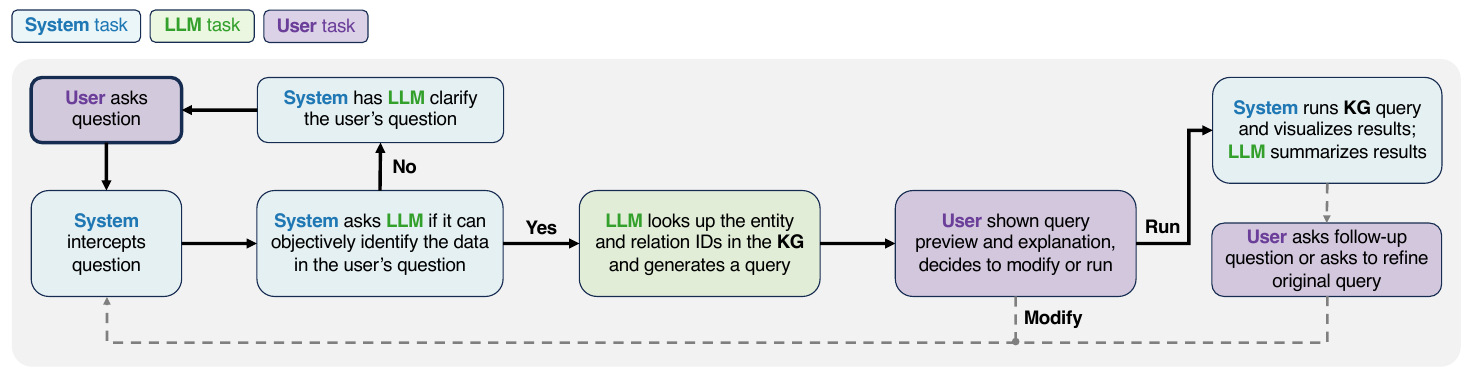}
    \caption{An overview of LinkQ's~\cite{li2024linkq} human-in-the-loop system design, described in Section~\ref{sec:design}. The \textbf{LLM}, \textbf{System}, and \textbf{User} have their own responsibilities for completing the question-to-query translation: the LLM constructs queries, the System works to illuminate possible hallucinations, and the User iterates with the System and LLM to modify, refine, or follow-up on generated queries.
    }
    \label{fig:pipeline}
\end{figure}

\sys uses a prompting protocol illustrated in Figure~\ref{fig:pipeline}, in which the LLM works with the System to query a KG based on the user's questions. The exact prompts and algorithmic protocols used for {\sys} are explained in more detail in LinkQ's original publication~\cite{li2024linkq} and are available open source at \url{https://github.com/mit-ll/linkq}.

First, the user and LLM engage in a back-and-forth dialogue to ensure a well-formed KG query can be written (Figure~\ref{fig:teaser}). To construct the query, the LLM and System work together to fuzzy search for entities (`nodes'), find relevant relations (`edges'), and traverse the KG to identify all the correct data IDs. Whenever the LLM needs to access data in the KG, LinkQ intercepts these requests and calls the appropriate KG API, then passes the KG graph structure and ground-truth IDs to the LLM via System messages.

Finally, the system provides the LLM with few-shot training~\cite{fewShotPaper} to write an accurate KG query. LinkQ visually displays the ground-truth query results in tabular format and provides an LLM-generated summary of the results. By having the LLM generate a query instead of directly answering the user's question, LinkQ both mitigates LLM hallucinations and retrieves up-to-date data for its answers. 
% \sys typically takes 10-15 seconds to complete the protocol of converting the User's natural language question to an LLM-generated KG query and query explanation, as demonstrated in Section~\ref{sec:quant-results}.
\section{Quantitative Evaluation of LinkQ}
\label{sec:results}

We conducted a quantitative evaluation to test LinkQ's capabilities in mitigating hallucinations by generating accurate queries from \textit{realistic} (i.e. not overly simplistic) natural language questions, using the Wikidata KG~\cite{wikidata2024stats}. We compared LinkQ to a standalone LLM (OpenAI's GPT-4) to understand when LinkQ's prompting strategy (Figure~\ref{fig:pipeline}) performs well or poorly. 
% This quantitative evaluation is distinct from testing \sys's user interface (Section~\ref{sec:interface-design}) which is evaluated in Sections~\ref{sec:demonstration} and~\ref{sec:bron-case-study}.
The entirety of our quantitative evaluation is available at \url{https://github.com/mit-ll/linkq}.

\subsection{Evaluation Design}

\begin{table}[h]
\centering
\small
\renewcommand{\arraystretch}{1.2}
\centering
% \vspace{-3mm}
\resizebox{.90\linewidth}{!}{%
\renewcommand\theadalign{bt}
\renewcommand\theadfont{\bfseries}
\renewcommand\theadgape{\Gape[4pt]}
\renewcommand\cellgape{\Gape[4pt]}
\setlength{\tabcolsep}{8pt}

\small
\sffamily
\begin{tabular}{lll}
\toprule
\textbf{Complexity Type} & \textbf{Topic} & \textbf{Exemplar Question}          
\\
\midrule 
Multi-hop         & books          & Where was the Nobel Prize in Literature winner from 2002 born?              \\
% \midrule 
Comparative       & sports       & Who has won more NBA Season MVPs, LeBron James or Steph Curry?                   \\
% \midrule 
Yes/No            & politics       & Was John Adams the second president of the United States?                   \\
% \midrule 
Generic           & history        & What year did Marie Curie win the Nobel Prize for Chemistry?                \\
% \midrule 
Intersection      & music          & Which member of the New Kids on the Block was the brother of Mark Wahlberg? \\ 
\bottomrule
\end{tabular}

}
\caption{A subset of questions that we tested LinkQ on, from the Mintaka~\cite{sen2022mintaka} \textit{complex questions} dataset.}
\label{tab:mintaka-questions}
% \vspace{-8mm}
\end{table}

% \noindent 
% \textbf{Dataset}:
% We selected the Mintaka~\cite{sen2022mintaka} Wikidata question bank, which provides a set of ``\textit{complex, natural questions}'' that are categorized into 9 question complexity types: count, comparative, superlative, ordinal, multi-hop, intersection, difference, yes/no, and generic. For each question type, there are \textbf{8} topics: history, music, books, movies, video games, politics, geography, and sports. 

\vspace{5pt}
\noindent 
\textbf{Dataset and Question Selection}:
\label{sec:eval-questions}
We selected a portion of questions from the Mintaka~\cite{sen2022mintaka} Wikidata question bank, focusing on \textbf{5} question types that we posited to be \textit{realistic} for natural language to KG query generation: (1) \textbf{Multi-hop}, (2) \textbf{Comparative}, (3) \textbf{Yes/No}, (4) \textbf{Generic}, and (5) \textbf{Intersection}. An example of each question type we tested is shown in Table~\ref{tab:mintaka-questions}. We randomly selected \textbf{3} questions for each of the \textbf{8} domain topics (history, music, etc.), ensuring all chosen questions adhered to these criteria: (1) the question can be objectively answered, (2) the answer is not dependent on out-of-date data, and (3) the answer can be currently found in Wikidata. For Yes/No questions, we selected an even split of ``yes'' and ``no'' answers. For Comparative questions, we did not select any that could result in a binary answer to avoid overlap with Yes/No questions. In total, our evaluation dataset has 120 questions (\textbf{5} types $\times$ \textbf{8} topics $\times$ \textbf{3} questions).

\vspace{5pt}
\noindent 
\textbf{Baseline}: We used off-the-shelf GPT-4 as the baseline method, instructing it to generate an appropriate KG query that finds the answer in Wikidata. Comparing the results of LinkQ (which uses GPT-4 under the hood, along with its message passing and prompting strategy) against plain GPT-4 provides evidence of how well LinkQ's prompting strategies improve upon the usage of a standalone LLM for KG query generation.

\vspace{5pt}
\noindent 
\textbf{Correctness Criteria}: We executed each of the questions from our curated dataset \textbf{three independent times} to account for the probabilistic nature of LLMs with non-zero temperatures~\cite{bender2021stochastic}. 
If LinkQ or GPT-4 produced a correct answer 
% -- from the query results or in its summary of the results -- 
\textit{at least once} from those three attempts, it was counted as a correct answer.
An incorrect answer could be due to a query that is syntactically incorrect, finds a wrong answer, 
% (e.g., due to hallucinated IDs), 
or returns empty results.
% (e.g., due to a misunderstanding of the KG schema). 

\subsection{Analysis of Results}
\label{sec:quant-results}

\begin{figure}[!h]
    \centering
    \subfigure[Question accuracy]{\includegraphics[width=0.525\textwidth]{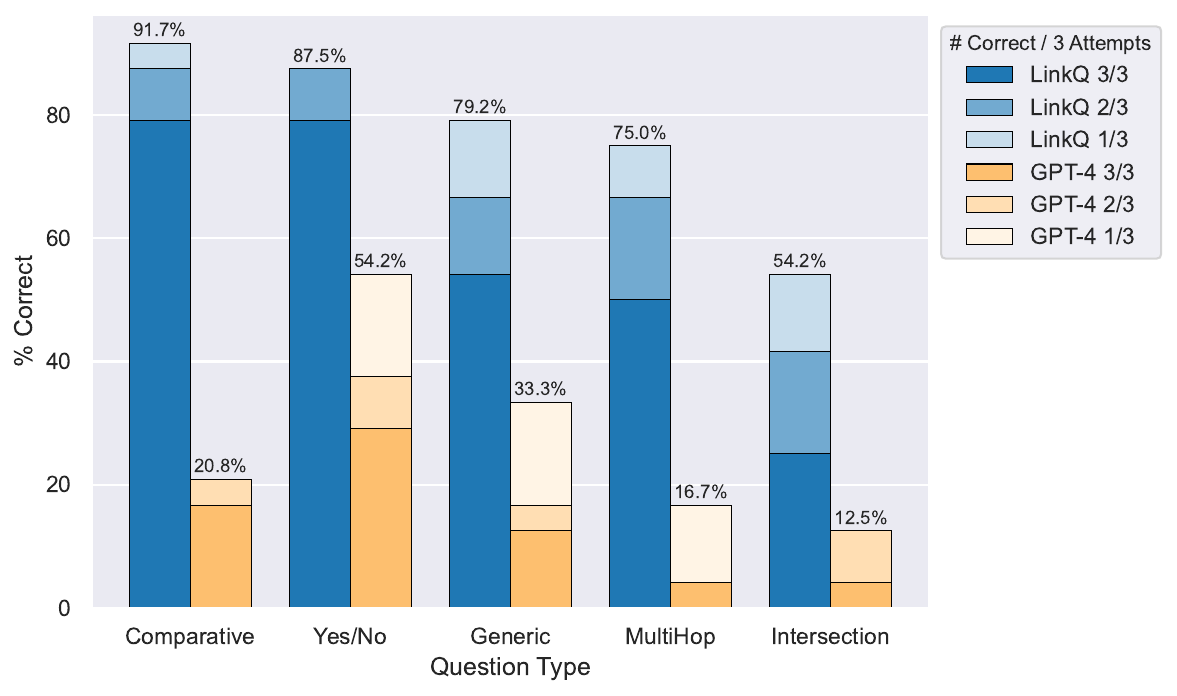}} 
    \subfigure[Runtime for query generation]{\includegraphics[width=0.4\textwidth]{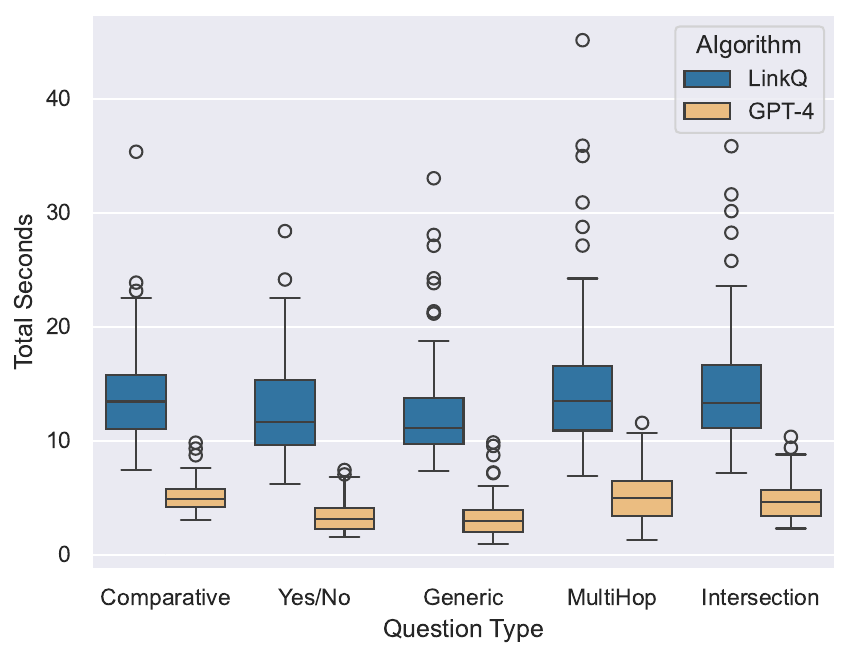}} 
    \caption{Results from our quantitative evaluation (Section~\ref{sec:results}) comparing LinkQ (blue) and GPT-4 (orange). 
    Left: LinkQ and GPT-4's overall question accuracy on 24 questions for each question type (x-axis), with a breakdown showing the number of correct attempts per question (y-axis). 
    Right: LinkQ versus GPT-4's runtime to generate a corresponding query.}
    \label{fig:eval-results}
    \vspace{-4.8mm}
\end{figure}

LinkQ outperformed GPT-4 in accuracy for every question type tested, indicating that LinkQ's message passing and prompting strategy increases the objective correctness (and consequently the runtime) of an LLM in translating natural language questions to KG queries without any fine-tuning. Figure~\ref{fig:eval-results} shows the full results. 
% Interestingly, there is negligible difference between LinkQ and GPT-4 in producing syntactically correct (i.e. error free) queries -- with LinkQ (99.2\%) slightly outperforming GPT-4 (97.5\%), and GPT-4 produces queries faster than LinkQ. 

The \textbf{Comparative} question type resulted in LinkQ's highest performance (91.7\%), while the \textbf{Yes/No} type resulted in GPT-4's (54.2\%). Since Yes/No questions were the only tested complexity type that could result in a binary true or false, it is possible that GPT-4's performance on this type is due to random chance -- since all other question categories resulted in far lower accuracy for GPT-4. 
In general, Comparative, Yes/No and Generic question types are common across knowledge graph question-answering (KGQA) sets, as they reflect low-level analysis questions that users would typically ask -- e.g., when using a natural language interface for initial data exploration. LinkQ's performance on these categories suggests that using its prompting strategy is sufficient when answering straightforward analysis questions.

The complexity of Multi-hop questions (which require at least two steps to answer correctly) is reflected in our evaluation results (LinkQ: 75\%, GPT-4: 16.7\%) as well as the runtime required to translate. Based on LinkQ's (54.2\%) and GPT-4's (12.5\%) performance on Intersection questions (which require pattern-matching on two or more conditions to answer correctly), they seem the most difficult for an LLM to correctly translate into KG queries. 

\smallbreak 
\noindent 
\textbf{Learning Outcomes:} 
Future systems should explore how breaking a complex question down into smaller steps---perhaps by following a chain-of-thought approach---could improve performance. Altogether, our results suggest that systems like LinkQ can perform well on writing KG queries that look for data existence, conduct fact-checking, and retrieve data from more than one to two hops away. However, certain complex questions that are Multi-hop, Intersection -- and possibly others we did not include in this evaluation -- will need further prompting strategies if avoiding fine-tuning. Ultimately, if the LLM is unsuccessful in translating a question into a KG query, the LLM should ask the user to help it transform the query step-by-step (i.e. follow LinkQ's human-in-the-loop approach) to avoid it hallucinating. 

% Because the LinkQ query building protocol involves iterative messages between the LLM and System as well as various KG API calls, it naturally is more time consuming than directly asking the LLM to generate a query without any context. LinkQ took on average 13.9 seconds to generate a query, versus GPT-4 taking 4.3 seconds, shown in Figure~\ref{fig:eval-results}.

% While the results of our quantitative evaluation highlight LinkQ's ability to generate successful KG queries -- without the burden of fine-tuning or KG embeddings -- LinkQ was designed to \textit{collaborate} with users for correct query construction, with supplemental tables and visualizations. We discuss how to leverage the learning outcomes of this quantitative evaluation, while maintaining \sys's original design goals, in Section~\ref{sec:future}.

\vspace{-1mm}
\section{Qualitative Study with Domain Experts}
\label{sec:demonstration}

% \begin{figure*}[!ht]% specify a combination of t, b, p, or h for top, bottom, on its own page, or here
%   \centering % avoid the use of \begin{center}...\end{center} and use \centering instead (more compact)
%   % trim={left bottom right top}
%   \includegraphics[clip, trim=0cm 4.7cm 0cm 3.25cm, width=.99\linewidth]{figures/linkq_bron_apt_techniques.pdf}
%   \caption{%
%   	An example use case of LinkQ integrated with a cybersecurity knowledge graph, BRON. During a demonstration, a subject matter expert in cybersecurity asked LinkQ, ``\textit{For all APTs (aka groups), what are the top 5 common techniques that they use? Also give me the technique names.}'' LinkQ correctly generated a SPARQL query that aggregated the top 5 most common techniques connected to groups in BRON, and included the count and technique names.
%   }
%   \label{fig:linkq_bron_group_techinques}
% \end{figure*}

% \begin{figure}[h!]
%     \includegraphics[width=0.65\textwidth]{figures/bron.pdf}
%   \caption{A visual schema of the cybersecurity knowledge graph, BRON, as illustrated by Hemberg et al.~\cite{hemberg2020linking} }
%   \label{fig:bron}
% \end{figure}

We discuss the results of a qualitative study with the BRON~\cite{hemberg2020linking} knowledge graph with two subject matter experts (SMEs). BRON is a cybersecurity graph database that connects MITRE's ATT\&CK MATRIX of Tactics and Techniques, NIST's Common Weakness Enumerations, NIST's Common Vulnerabilities and Exposures, and NIST's Common Attack Pattern Enumeration and Classification list. While these data sources on their own are informative, they are highly dependent on one another. Therefore, BRON---a unification of these independent data sources---provides necessary context to cybersecurity experts who investigate the linkage of cyber alerts, threats, and vulnerabilities~\cite{milajerdi2019poirot}. 
At the time of this writing, our tests suggest GPT-4 has not been trained on BRON, which could result in it hallucinating if SMEs wanted to ask questions about BRON without LinkQ. 

\noindent 
\vspace{2pt}
\noindent 
\textbf{LinkQ Demonstration}: { 
As BRON is extremely domain-specific, we worked with two subject matter experts in cybersecurity who are end-users of BRON. These SMEs do not consider themselves KG experts; regardless, they must use BRON in their own daily work.
Both SMEs sought to test whether LinkQ could be used to support their previous work with BRON. The first question they asked LinkQ was: ``\textit{Give me a list of all available APTs [advanced persistent threats],}'' an entity type in BRON that represents cyberattacker groups, ``\textit{and all of their properties}.''
%SMEs told us that this list is necessary from BRON because cyberattacks and relationships between cyberattacks are often updated.
LinkQ successfully wrote and executed the query to retrieve all APTs and their properties.  From this list, the SMEs had LinkQ (successfully) retrieve all possible \textit{techniques} associated with a specific APT, which would inform them of the methodologies behind a cyberattack of interest. Finally, we asked the SMEs to think of an example of a particularly difficult question they had tried to answer previously with BRON. The question was a method of aggregating BRON data: ``\textit{For all the APTs, what are the top five most common techniques used among all of them?}''  Both SMEs were surprised to see that LinkQ generated the answer to their question. See Figure~\ref{fig:teaser} for an example of querying BRON.}

% \item 
\vspace{2pt}
\noindent 
\textbf{Learning Outcomes}: { We find that LinkQ is successful, but not 100\%, at querying the domain-specific KG BRON. At the start of our live demonstration, both SMEs were skeptical of LinkQ being capable of extracting specific relationships in BRON, as they believed BRON was too complex to for an LLM to navigate.
One SME described the difficulty they faced for writing queries to identify all possible paths in BRON between multiple node types (e.g., finding techniques that are associated with APTs -- as tested in our demonstration), as it requires an in-depth understanding of the BRON schema. This finding highlights the ability of LinkQ to support question-answering over domain-specific KGs, thus addressing an LLM's tendency to hallucinate when it does not know the answer to domain-specific questions.

Throughout the demonstration, LinkQ did not hallucinate false query results but at times stated there was no associated data with the user's question (i.e. the query would return no results) -- when in fact the data did exist. Empty results most often occurred when the LLM did not thoroughly explore the KG to capture the needed structure of the query. Since the SMEs were familiar with the BRON KG, they were often able to iterate with LinkQ to correct the query (e.g., by specifying the correct relationships); of course, this is a more difficult task if the end-user is not an expert with the KG. Along with this limitation, both SMEs had a number of suggestions for improving LinkQ for their use cases: 
% (1) give users the ability to edit the initial LLM instructions so they can specify what data they care about, (2) allow users to query for entities based on their written descriptions, rather than their properties, (3) have the LLM generate a ranked list of potential queries that the user can choose from, and (4) provide additional semantic context for entities or properties in BRON from external data sources.

\begin{itemize}[topsep=0pt, partopsep=0pt,itemsep=0pt,parsep=0pt,leftmargin=*]
\item \textbf{Give users the ability to edit the initial LLM instructions so they can specify what data they care about.} SMEs noted that the LLM would retrieve data they knew was irrelevant for their task. They suggested it would be helpful if they could edit the initial system prompt to tell the LLM which data to ignore or highlight.

\item \textbf{Allow users to query for entities based on their written descriptions, rather than their properties.} One SME wanted the LLM to query BRON based on the descriptions for nodes contained in the graph database, rather than searching based on names. In this case, a RAG~\cite{rag} approach might be preferred over LinkQ's approach to Q\&A.
% \strike{ Future work can investigate translating user's descriptions for entities into KG queries.}

\item \textbf{Determine a ranked list of potential queries (or query results) when the LLM is uncertain.} Both SMEs told us they wanted to see multiple query versions of their question so they could determine which one to execute, especially when the LLM may have translated it incorrectly. 
% \strike{This feature could be computationally expensive to run with an LLM, as it might require multiple API calls.} 
Future work will need to investigate the cost-to-benefit ratio of this, or offer it as an optional feature that can be selected or deselected by the end-user. 
% \strike{Moreover, future work will need to seek to understand the uncertainty of an LLM's generated query beyond the LLM's own knowledge of the data and query syntax, which can help prevent unnecessary query generation.}

\item \textbf{Provide extra semantic context for entities or properties in the KG from external data sources.} SMEs noted that missing data about entities in BRON could be found from other cyber frameworks and suggested that it would be beneficial if future systems could supplement its query results with additional context from external sources. 

\end{itemize}

\section{Conclusion}

This workshop paper evaluated the efficacy of using LinkQ to mitigate LLM hallucinations by generating knowledge graph queries that retrieve up-to-date, ground-truth data. Our quantitative evaluation showed that \sys outperforms GPT-4 in accurately answering questions through the construction of KG queries, although it still struggles with certain complex question types. From a qualitative study with practitioners on a real-world cybersecurity KG, we found that \sys exceeded expectations for converting NL questions to KG queries, supporting them in answering domain-specific questions that an LLM would otherwise not know the answers to. We presented concrete opportunities for improving future systems like LinkQ that still leverage its grounded question-answering capabilities while addressing additional real-world needs. Overall, our evaluations of LinkQ suggest that LLMs can be adapted for high-stakes domains when designed with appropriate guardrails to combat hallucinations so they can be used responsibly.

\begin{acks}
We thank our collaborators at MIT Lincoln Laboratory for participating in our qualitative study and offering their feedback on LinkQ to shape future use cases. We also thank the reviewers for their time and thoughtful suggestions to help improve this workshop paper.
\smallbreak 

\noindent
% \small
DISTRIBUTION STATEMENT A. Approved for public release. Distribution is unlimited. This material is based upon work supported by the Combatant Commands under Air Force Contract No. FA8702-15-D-0001. Any opinions, findings, conclusions or recommendations expressed in this material are those of the author(s) and do not necessarily reflect the views of the Combatant Commands. © 2024 Massachusetts Institute of Technology. Delivered to the U.S. Government with Unlimited Rights, as defined in DFARS Part 252.227-7013 or 7014 (Feb 2014). Notwithstanding any copyright notice, U.S. Government rights in this work are defined by DFARS 252.227-7013 or DFARS 252.227-7014 as detailed above. Use of this work other than as specifically authorized by the U.S. Government may violate any copyrights that exist in this work.

\smallbreak 
\noindent This work was authored in part by the National Renewable Energy Laboratory (NREL), operated by Alliance for Sustainable Energy, LLC, for the U.S. Department of Energy (DOE) under Contract No. DE-AC36-08GO28308. The views expressed in the article do not necessarily represent the views of the DOE or the U.S. Government. The U.S. Government retains and the publisher, by accepting the article for publication, acknowledges that the U.S. Government retains a nonexclusive, paid-up, irrevocable, worldwide license to publish or reproduce the published form of this work, or allow others to do so, for U.S. Government purposes.
\normalsize
\end{acks}

\bibliographystyle{CHI-ACM-contents/ACM-Reference-Format}
\bibliography{main}

\appendix

\end{document}